\documentclass[sigconf,natbib=true]{acmart}

\AtBeginDocument{%
  \providecommand\BibTeX{{%
    \normalfont B\kern-0.5em{\scshape i\kern-0.25em b}\kern-0.8em\TeX}}}

\setcopyright{acmcopyright}
\copyrightyear{2018}
\acmYear{2018}
\acmDOI{XXXXXXX.XXXXXXX}

\acmConference[Conference acronym 'XX]{Make sure to enter the correct
  conference title from your rights confirmation emai}{June 03--05,
  2018}{Woodstock, NY}
\acmPrice{15.00}
\acmISBN{978-1-4503-XXXX-X/18/06}

\newcommand{\ie}{\emph{i.e., }}
\newcommand{\eg}{\emph{e.g., }}

\newcommand{\wrt}{\emph{w.r.t. }}
\newcommand{\cf}{\emph{cf. }}

\usepackage{enumitem}
\usepackage{verbatim}
\usepackage[ruled]{algorithm2e}
\usepackage{multirow}
\usepackage{subfigure} 
\usepackage{ragged2e}
\usepackage{caption}
\usepackage{graphicx}
\usepackage[normalem]{ulem}
\useunder{\uline}{\ul}{}
\usepackage{amsmath}
\usepackage{setspace}
\usepackage{comment}
\usepackage{bm}
\usepackage{color}
\usepackage{wrapfig}

\begin{document}

\fancyhead{}
\title{Understanding and Counteracting Feature-Level Bias in Click-Through Rate Prediction}

\author{Jinqiu Jin$^{1}$, Sihao Ding$^{1}$, Wenjie Wang$^{2}$, Fuli Feng$^{1\dag}$}

\def\authors{Jinqiu Jin, Sihao Ding, Wenjie Wang, Fuli Feng}

\affiliation{\institution{$^1$University of Science and Technology of China,$^2$National University of Singapore\country{}}}

\email{{jjq20021015,dsihao}@mail.ustc.edu.cn, {wenjiewang96,fulifeng93}@gmail.com}

\thanks{$\dag$ Corresponding author.}

\begin{abstract}
  Common click-through rate (CTR) prediction recommender models tend to exhibit feature-level bias, which leads to unfair recommendations among item groups and inaccurate recommendations for users. While existing methods address this issue by adjusting the learning of CTR models, such as through additional optimization objectives, they fail to consider how the bias is caused within these models. To address this research gap, our study performs a top-down analysis on representative CTR models. Through blocking different components of a trained CTR model one by one, we identify the key contribution of the linear component to feature-level bias. We conduct a theoretical analysis of the learning process for the weights in the linear component, revealing how group-wise properties of training data influence them. Our experimental and statistical analyses demonstrate a strong correlation between imbalanced positive sample ratios across item groups and feature-level bias. Based on this understanding, we propose a minimally invasive yet effective strategy to counteract feature-level bias in CTR models by removing the biased linear weights from trained models. Additionally, we present a linear weight adjusting strategy that requires fewer random exposure records than relevant debiasing methods. The superiority of our proposed strategies are validated through extensive experiments on three real-world datasets.
\end{abstract}



\maketitle
\section{Introduction} \label{sec:intro}

\begin{figure}
    \centering
    \includegraphics[width=0.99\linewidth]{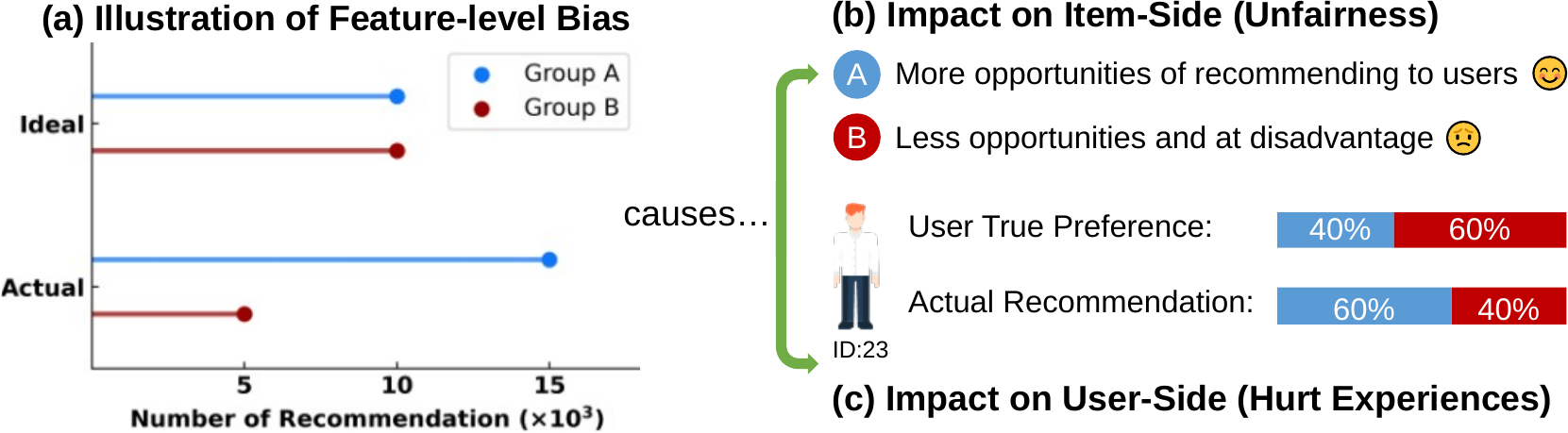}
    \vspace{-5pt}
    \caption{{A toy example to illustrate feature-level bias. (a) \textit{Ideal} and \textit{Actual} denote the expected number of recommendation by ground truth and by CTR model, respectively. (b) Item-side unfairness, \textit{group A} and \textit{group B} are over-recommended and under-recommended, respectively. (c) Hurting user experiences, the model recommends 60\% A and 40\% B to a user (\textit{ID=23}), whereas his true preference consists of 40\% A and 60\% B.}}
    \vspace{-10pt}
    \label{fig:intro}
\end{figure}

Recommender systems are commonly used to address information overload on various web platforms by providing personalized information retrieval~\cite{ncf,deepfm}. Click-through-rate prediction is a critical task in recommender systems that leverages user-item features to rank candidate items for better personalization~\cite{huawei_BARS}. However, CTR models are susceptible to \textit{feature-level bias}, which leads to over-recommendation or under-recommendation of different item groups based on a feature field (e.g., genre). This bias can cause unfairness among item groups and negatively impact user experiences. Figure~\ref{fig:intro} illustrates this issue through a toy example where \textit{group A} is over-recommended while \textit{group B} is under-recommended, leading to deviation from the true preference distribution of some users. It is thus essential to counteract the feature-level bias of CTR models.

Existing methods mainly deal with the feature-level bias by adjusting the learning of CTR models. For instance, a) regularization~\cite{geyik2019fairness}, which adjusts the learning objective by applying fairness or bias oriented terms to regulate the outputs of CTR models; b) adversarial training~\cite{zhu2020measuring}, which adjusts the learning procedure by adding an adversary to control the discrimination between groups at the level of item representations or model outputs; c) causal inference~\cite{decrs}, which adjusts the learning target as the causal effect of exposure through the potential outcome framework~\cite{rubin2005causal} or structural causal models~\cite{pearl2009causality}. Nevertheless, there remains an open problem about the mechanism of leading to the feature-level bias in CTR models, \ie how the bias generates from raw training data to biased model outputs, which is important to the development of trustworthy CTR models and the regulation of CTR models~\cite{trecsurvey}. This calls for answering the following question:
\textit{Which component of the CTR model mostly explains the feature-level bias?}

Considering the wide usage of Factorization Machine~\cite{rendle_fm} (FM) and its extensions~\cite{nfm,afm,deepfm}, we first study the role of linear feature projection and feature interaction in misleading the recommendation. 
Specifically, we compare the recommendations of well-trained FM and NFM~\cite{nfm} when their linear component or high-order component is blocked. Results on benchmark datasets reveal the dominant role of the linear component in differentiating item groups, \ie generating feature-level bias. Subsequently, we conduct theoretical analysis on the learning process of the linear weights related to the bias field, discovering their connection to the statistics of positive and negative samples in the training data. Lastly, results of statistical analysis further validate that the uneven distribution of positive sample ratio over item groups with different feature values is the cause of feature-level bias.

Taking one step further, according to the analysis, we pursue minimally invasive strategies for counteracting the feature-level bias to avoid the overhead of adjusting the learning of CTR models. In particular, we design two debiasing strategies that perform minimally invasive treatments on the well-trained CTR model by adjusting the linear weights. To block the impact of the feature-level bias, we first propose a linear weight reduction strategy, which controls the contribution of linear weights with a linear coefficient. Furthermore, we pursue better estimation of user preference by adjusting the linear weights according to limited random exposure data with a linear weight reconstruction strategy. We conduct extensive experiments on three benchmark datasets. The results show the effectiveness of the strategies, which can remarkably improve item-side fairness and recommendation performance. The code and data are available at \url{https://github.com/mitao-cat/feature-level_bias.git}.

In a nutshell, the contributions of this work are three folds:
\begin{itemize}
    \item We uncover the mechanism of generating feature-level bias in CTR models, which reveals the key role of linear weights and the positive sample ratio.
    \item We design two simple-yet-effective methods for counteracting feature-level bias, which is applicable to widely-used FM-based CTR models with limited overhead.
    \item We conduct extensive experiments on three benchmark datasets, validating the rationality and effectiveness of our methods.
\end{itemize}

\section{PRELIMINARIES}
In this section, we briefly introduce the basic concepts of CTR prediction and feature-level bias.
\subsection{CTR Prediction}
CTR prediction is an important task in recommender systems, which aims at using abundant features to estimate the probability of user clicks over items~\cite{huawei_BARS}. The key of the CTR task lies in the comprehensive modeling of the rich user and item features. Specifically, we introduce the CTR prediction task as follows. 

\textbf{Input.} 
We denote the training samples for CTR models as $\mathcal{D}=\{(\bm{x},y)\}$, where $\bm{x}\in\mathbb{R}^n$ is a sparse vector and $y\in\{0,1\}$ is the label (\ie click or not). The label is typically collected from previous exposure strategies. The feature vector describes a pair of user and item with $n$ features. Following the previous work ~\cite{FFM}, we suppose that the feature vector $\bm{x}$ contains different \textit{fields}, where a field contains different \textit{features} corresponding to different categories of the field. For example, we represent the click of a male user on an action and sci-fi movie as:
\begin{equation}
    \bm{x}=[\underbrace{0.5,0.5,0,\cdots,0}_{\text{\textit{Genre=\{Action,Sci-Fi\}}}}, \underbrace{0,\cdots,0,1,0}_{\text{\textit{MovieID=1024}}}, \underbrace{\cdots}_{\cdots}, \underbrace{0,1}_{\text{\textit{Gender=Male}}}]
\end{equation}
where \textit{Genre}, \textit{MovieID} and \textit{Gender} are fields with values \textit{\{Action,Sci-Fi\}}\footnote{Note that some fields may take multiple feature values, \ie a user or item could belong to different groups at the same time.}, \textit{Male} and \textit{1024}, respectively. 

\textbf{Prediction.} The CTR model typically estimates the click probability $P(y=1|\bm{x})$ with consideration of both raw features and feature-interactions. We focus on the classical FM~\cite{rendle_fm} and its following extensions which are widely used in practical applications. These FM-based CTR models can be abstracted as:
\begin{equation}\label{eq:fm_form}
     \hat{y} = w_0+\sum_{i=1}^nw_ix_i+f_{\theta}(\bm{x}),
\end{equation}
where $w_0$ is the global bias\footnote{Note that $w_0$ is omitted in the following analysis since it is constant to items with different feature values.}; $\sum_{i=1}^nw_ix_i$ is a linear component where $w_i$ models the direct contribution of the $i$-th feature to the prediction of clicks; and $f_{\theta}(\bm{x})$ with parameters $\theta$ models feature interactions which has model-specific design, \eg FM~\cite{rendle_fm} uses feature embeddings to model second-order interactions, NFM~\cite{nfm} uses deep neural networks to model higher-order interactions.

\textbf{Optimization.} The parameters of a CTR model is optimized over the historical clicks $\mathcal{D}$ by minimizing a recommendation loss, such as the commonly used binary cross-entropy (BCE) loss:
\begin{equation}
    \mathcal{L}=-\frac{1}{|\mathcal{D}|}\sum\limits_{(\bm{x},y)\in\mathcal{D}}\left(y\log \sigma\left(\hat{y}\right)+\left(1-y\right)\log\left(1-\sigma\left(\hat{y}\right)\right)\right),
\end{equation}
where $|\mathcal{D}|$ denotes the number of training samples and $\sigma(\cdot)$ denotes the \textit{sigmoid} function to normalize the prediction. For briefness, we omit the regularization term such as the $L_2$-norm, which is widely used to prevent overfitting.

\subsection{Feature-level Bias}
Feature-level bias in CTR prediction refers that \textit{CTR models may over/under-recommend different item groups \wrt a given feature field.} In other words, the recommendation is biased towards some item groups which are given more recommendation opportunities than deserved~\cite{hardt2016equality}, \ie some items are over-scored due to their bias feature. Given a CTR model, we study the feature-level bias of its recommendations (one ranking list per user) over a set of testing samples $\mathcal{D}_t$. As to user $u$, let $k_u^+$ and $k_u^-$ denote the number of positive and negative samples in the testing set $\mathcal{D}_t$ given by the user. Accordingly, we take the top-$k_u^+$ ranked items in the ranking list of user $u$ as the exposures, which is denoted as $\mathcal{P}_u$. Without loss of generality, we assume the \textit{bias field} has $k$ features, \eg there are $k$ different video tags, leading to $k$ item groups. Furthermore, we assume the bias field is placed at the beginning of the feature vector $\bm{x}$, \ie $\{x_j | j \in [1, k]\}$ are the bias features. In particular, we adopt the following metric to measure the feature-level bias:
\begin{itemize}[leftmargin=*]
    \item \textit{Exposure-to-Hit Ratio} ($EHR$).  Given a feature $j$ in the bias field, the \textit{Exposure-to-Hit Ratio} can be defined as:
    \begin{equation}\label{eq:ratio}
        EHR(j)=\frac{\sum_u \sum_{(\bm{x}, y) \in \mathcal{P}_u} I(\bm{x}, j)}{\sum_{(\bm{x},y) \in \mathcal{D}_t} I(\bm{x}, j) \land y },\quad
        I(\bm{x}, j) =
        \begin{cases}
            1 &\mbox{$x_j > 0$,}\\
            0 &\mbox{else.}
        \end{cases}
    \end{equation}
    $I(\bm{x}, j)$ indicates whether the sample has the bias feature, \ie the related item belongs to the $j$-th group. $I(\bm{x}, j) \land y=1$ means a user click on the $j$-th group ($\ie I(\bm{x}, j) = 1$), $I(\bm{x}, j) = 0$ otherwise. 
\end{itemize}

\begin{figure}
    \centering
    \includegraphics[width=0.99\linewidth]{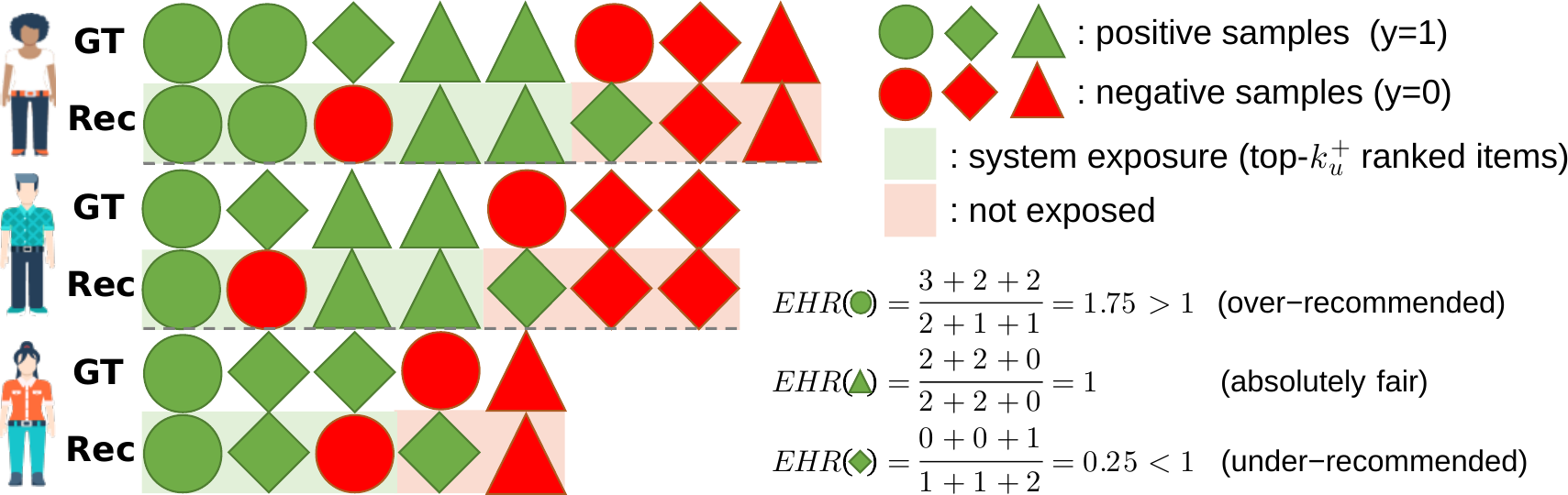}
    \vspace{-5pt}
    \caption{An example to illustrate the calculation of EHR. Circle, diamond and triangle denote three different item groups. \textit{GT} denotes the ground-truth of user interactions. \textit{Rec} denotes the ranked user list given by the CTR model.}
    \label{fig:EHR}
    \vspace{-10pt}
\end{figure}

$EHR(j)$ calculates the ratio between the \textit{number of exposures} (\ie ranked in user's top-$k_u^+$ list) belonging to the $j$-th group and the \textit{number of clicks} (\ie positive sample in user list) on the $j$-th group within the testing set $\mathcal{D}_t$. Figure~\ref{fig:EHR} gives an example about the calculation process of EHR. For the \textit{triangle} group, the model correctly ranks all positive samples in each user's top-$k_u^+$ list, without falsely putting forward any negative samples. Accordingly, $EHR(\triangle)=1$, which means the CTR model exactly gives the triangle group recommendation opportunities it deserves. In contrast, the CTR model shows strong bias towards the \textit{circle} group as compared to the \textit{diamond} group as the model tends to rank negative circle samples before the positive diamond samples. As a result, $EHR(\circ)>1$ (\ie over-recommended) and $EHR(\diamond)<1$ (\ie under-recommended), indicating their unequal recommendation opportunities~\cite{geyik2019fairness}.

\subsection{Further Discussion}

\subsubsection{Connection between EHR and Fairness Notions}

Feature-level bias has strong connection with item-side group fairness, which requires that different item groups should be treated similarly by the model~\cite{caton2020fairness}. Among the fairness notions, \textit{Equal Opportunity} (EO)~\cite{hardt2016equality} is the most related to Exposure-to-Hit Ratio. EO requires that the \textit{True Positive Rate} (TPR) across groups are the same~\cite{geyik2019fairness}. Formally, 
\begin{equation}\label{eq:tpr}
    TPR(j)=\frac{\sum_u \sum_{(\bm{x}, y) \in \mathcal{P}_u} I(\bm{x}, j) \land y}{\sum_{(\bm{x},y) \in \mathcal{D}_t} I(\bm{x}, j) \land y }=c,\ j\in[1,k],
\end{equation}
where $c\in[0,1]$ denotes a constant value. However, as TPR takes value between 0 and 1, it's hard to judge whether the item group is over-recommended (\ie at advantage) or under-recommended (\ie at disadvantage) by the CTR model, which is the key of analyzing feature-level bias. In contrast, the Exposure-to-Hit Ratio reflects the recommendation opportunity of an item group (see Figure~\ref{fig:EHR}). Comparing the EHR value to 1 allows for easy determination of whether a group is over-recommended or under-recommended. Furthermore, in the ideal case without feature-level bias (\ie $\forall j \in [1, k], EHR(j) = 1$), recommendation probabilities for different groups would be relatively similar and approach the EO standard. In this paper, we leverage Exposure-to-Hit Ratio to study the feature-level bias, and use conventional fairness notions (\eg EO) to measure its impact on item-side fairness.

\begin{figure}
    \centering
    \subfigure[ML-1M]{\includegraphics[width=0.5\linewidth]{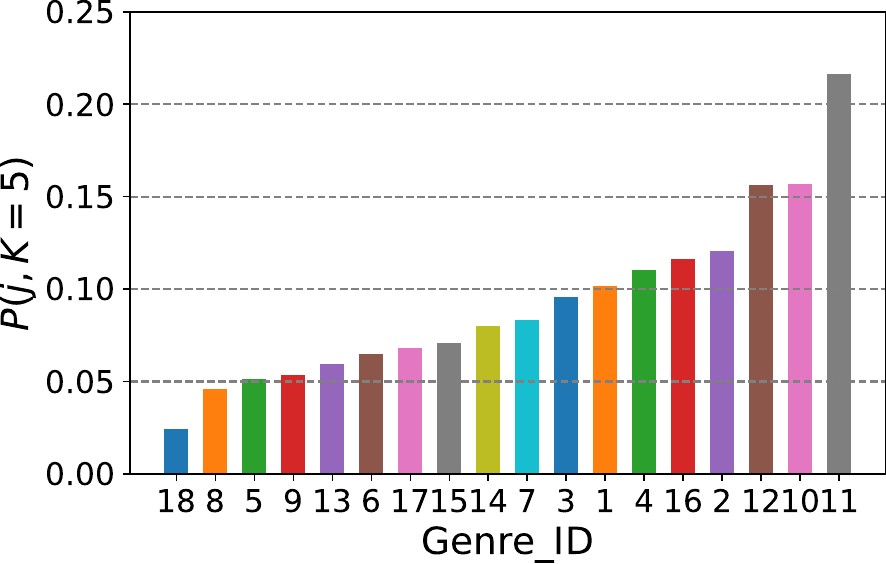}\label{subfig:eo5}}\qquad
    \subfigure[KuaiRand]{\includegraphics[width=0.4\linewidth]{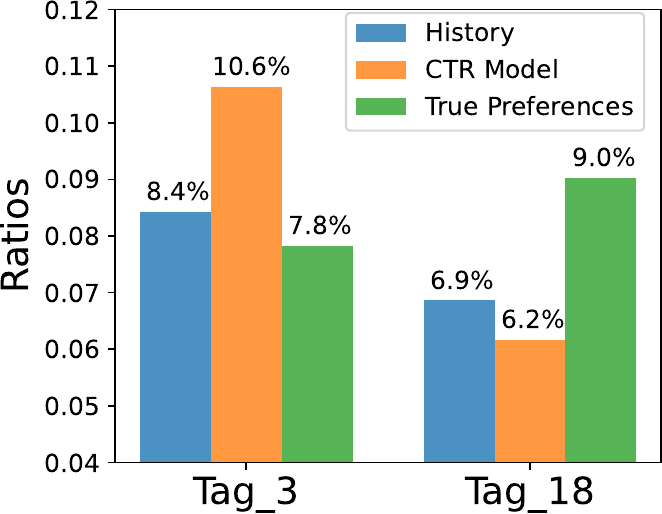}\label{subfig:biasamp}}
    \vspace{-10pt}
    \caption{(a) Illustration of item-side unfairness \wrt $P(j,K=5)$. (b) An example of the gap between \textit{History} and \textit{True Preferences} and the amplification effect of the CTR model. \textit{History} denotes the ratio of positive samples in a group to the total positive samples collected from previous recommendations. \textit{CTR Model} denotes the ratio of predicted positive samples in a group to the total positive samples collected from previous recommendations. \textit{True Preferences} denotes the ratio of positive samples in a group to the total positive samples collected from the random exposure.}
    \vspace{-10pt}
    \label{fig:bias_ctr}
\end{figure}

\subsubsection{Connection between Feature-level Bias and Recommendation Qualities}\label{sec:2.3.2}
Recent literatures~\cite{li2022fairness,chen2020bias} reveal that biases could produce unfair recommendation results and hurt user experiences. As for item-side, the items ranked in top places usually catch more user's attention~\cite{moffat2008rank}. If we regard the clicked items that are ranked in top-$K$ list of user $u$ (denoted as $\mathcal{P}_u^K$) as successful recommendations, by substituting $\mathcal{P}_u^K$ for $\mathcal{P}_u$ in Equation~\eqref{eq:tpr}, we obtain the new TPR values (denoted as $P(j,K)$) under the top-$K$'s consideration. Figure~\ref{subfig:eo5} illustrates the item-side unfairness issue with an example in movie recommendation, where $P(j,K)$ values vary greatly across item groups \wrt different movie genres, indicating the unequal opportunity of being recommended. As for user-side, the over-recommendation of some groups will amplify the impact of the gap between the user preference observed from the historical interactions and the true user preference~\cite{feedback_loop} as shown in Figure~\ref{subfig:biasamp}, which might hurt the user experience.

\section{MODEL ANALYSIS} \label{sec:analysis}

This section presents a top-down analysis of feature-level bias generation from the perspective of CTR models. We focus on two widely-used models, Factorization Machine (FM)~\cite{rendle_fm} and Neural Factorization Machine (NFM)~\cite{nfm}, as they are commonly used in both industry and academia. The analysis of more complex CTR models (\eg DIN~\cite{DIN}) is left for future work.

\subsection{Bias Analysis from Model View}

Firstly, we identify which component of the CTR model results in the feature-level bias in the recommendations. Functionally speaking, the core operations of CTR models are feature projection and feature interaction modeling. Accordingly, CTR models mainly consist of two parts: a) the \textit{high-order part} $f_{\theta}(\bm{x})$ in Equation~\eqref{eq:fm_form} for feature interaction, and b) the \textit{linear part} $\sum_{i=1}^nw_ix_i$ in Equation~\eqref{eq:fm_form} for feature projection. Thereafter, we aim to answer the question: \textit{to what extent the high-order and linear parts contribute to the feature-level bias in the output of CTR models?}

To answer the question, the key lies in studying how the two parts affect the biased distribution of prediction scores across item groups. To this end, we calculate the average prediction scores of the linear and high-order parts in each group, respectively; and then obtain the variance of scores across groups. The variance reflects the strength of two parts biasing the prediction scores across groups, where a larger variance indicates the stronger effect of injecting feature-level bias into prediction scores. In other words, the variance reflects to what extent the high-order or linear parts treat item groups differently. Since the testing data contains positive and negative samples, we need to calculate the variance separately to block the influence of sample labels. Formally, we compute the variance of the high-order part on positive samples as follows:
\begin{equation}\label{eq:variance}
    \mathcal{V}\left(\{m_j\}_{j=1}^k\right), \text{ where } 
    m_j = \mathop{\mathcal{M}}\left(\left\{f(\bm{x})|(\bm{x},y)\in\mathcal{D}_t, x_j>0,y=1\right\}\right),
\end{equation}
where $m_j$ is the average prediction score of the positive samples (\ie $y=1$) in the $j$-th group with feature $j$ (\ie $x_j >0$); $\mathcal{M}(\cdot)$ and $\mathcal{V}(\cdot)$ calculates the mean and variance, respectively. Similarly, to calculate the variance of the linear part \wrt positive samples, we only need to substitute $\sum_{i=1}^n w_i x_i$ for $f(\bm{x})$. Besides, calculating the variance on negative samples only needs to change $y=1$ with $y=0$ in Equation~\eqref{eq:variance}. 

\begin{table}[t]
\caption{Variance of the linear and high-order parts, where a larger variance implies a larger influence on the feature-level bias. "Positive" and "Negative" denotes that the variance is calculated over all positive or negative samples in the testing set, respectively.}
\label{tab:decompose}
\vspace{-10pt}
\resizebox{0.99\linewidth}{!}{
\begin{tabular}{lcccc}
\toprule
\multirow{2}{*}{\textbf{Model Part}} & \multicolumn{2}{c}{\textbf{FM}} & \multicolumn{2}{c}{\textbf{NFM}} \\
 & \textbf{Linear} & \textbf{High-order} & \textbf{Linear} & \textbf{High-order}\\ \hline 
$\mathbf{Positive}$ & 0.0582 & 0.0213 & 0.1026 & 0.0016 \\
$\mathbf{Negative}$ & 0.1233 & 0.0168 & 0.1572 & 0.0129 \\
\bottomrule
\end{tabular}}
\vspace{-10pt}
\end{table}

To compare the contributions of the linear and high-order parts to biased prediction scores, we train two representative CTR models, FM and NFM, on a popular CTR dataset ML-1M~\cite{movielens}. Thereafter, we calculate the variance of the two parts via Equation~\eqref{eq:variance}. The results are presented in Table~\ref{tab:decompose}, from which we find that the variance of the linear part is significantly higher than that of the high-order part. This indicates that \textit{the linear part is a critical reason for the biased prediction scores across item groups}, further leading to the feature-level bias in the recommendations. This is consistent with the function of two parts: the linear part models the strength of each feature through linear weights~\cite{rendle_fm}, thus the logit value may be biased towards certain feature groups; by contrast, as the high-order part models the complex interaction between multiple features, it should be less biased towards single feature groups.

\subsection{Bias Analysis from Data View}

Next, we further explore what causes the bias in the linear part from the data view. In the linear part, the linear weights $\{w_j|j\in\{1,\cdots,k\}\}$ in Equation~\eqref{eq:fm_form} are the only learnable parameters. A higher value of the linear weight $w_j$ is more likely to increase the final prediction scores of the samples with feature $j$, leading to the feature-level bias. As model weights are typically initialized with random values and learned in a data-driven manner, we postulate that the scale of the linear weight is related to the property of the training data. To discover how the data affect the linear weights during training, we analyze the relations between $w_j$ and a training sample $(\bm{x},y)$ by considering the gradient update. As the BCE loss for sample $(\bm{x},y)$ can be written as
\begin{equation}
    l=-y\log \sigma(\hat{y})-\left(1-y\right)\log\left(1-\sigma(\hat{y})\right)
    \label{eq:lj}.
\end{equation}
We can take the partial derivative of $l$ \wrt $w_j$ as follows:
\begin{equation}
\frac{\partial l}{\partial w_j} = \frac{\partial l}{\partial\hat{y}}\cdot\frac{\partial\hat{y}}{\partial w_j} = \left(-y(1-\sigma(\hat{y}))+(1-y)\sigma(\hat{y})\right)\cdot x_j = \left(\sigma(\hat{y})-y\right)\cdot x_j.
\end{equation}
By backward propagation with the learning rate $\epsilon$, the updated weight $w_j'$ should be:
\begin{equation}
    w_j'=w_j-\epsilon\frac{\partial l}{\partial w_j}= w_j+\epsilon\left(y-\sigma(\hat{y})\right)\cdot x_j,
\label{eq:update}
\end{equation}
where $\sigma(\hat{y})\in(0,1)$ is the prediction score for $\bm{x}$. From Equation~\eqref{eq:update}, we deduce that 1) a negative sample with the $j$-th feature (\ie $x_j>0$ and $y=0$) will decrease $w_j$ due to $y-\sigma(\hat{y})<0$; and 2) a positive sample with the $j$-th feature will increase $w_j$ in contrary since $y-\sigma(\hat{y})>0$. As such, a reasonable judgement is that \textit{the linear weights should be correlated to the number of positive and negative samples in the training set}, establishing the relations between the biased CTR model and the training data~\cite{chen2020bias}.

\begin{table}[t]
\caption{Correlation coefficients between model's linear weights and the three group-level dataset statistics $\mathbf{N_{p}}$, $\mathbf{N_{p}-N_{n}}$, and $\mathbf{N_{p}/(N_{p}+N_{n})}$. $SP$ and $PS$ denote the Spearman and Pearson coefficients, respectively, attached with $p$-values in the bracket. A larger absolute value of either SP or PS implies stronger correlation.}
\label{tab:twotest}
\vspace{-10pt}
\resizebox{0.99\linewidth}{!}{
\begin{tabular}{ccccc}
\toprule
\multicolumn{2}{c}{\textbf{Statistics}} & $\mathbf{N_{p}}$ & $\mathbf{N_{p}-N_{n}}$ & $\mathbf{N_{p}/(N_{p}+N_{n})}$ \\ \hline
 & $\mathbf{SP}$ & -0.2797 (3e-1) &  0.7110 (9e-4) & 0.8782 (2e-6) \\
\multirow{-2}{*}{\textbf{FM}} & $\mathbf{PS}$ & -0.0670 (8e-1) & 0.4086 (9e-2) & 0.8069 (5e-5) \\ \hline
 & $\mathbf{SP}$ & -0.4138 (9e-2) & 0.6615 (3e-3) & 0.8989 (4e-7) \\
\multirow{-2}{*}{\textbf{NFM}} & $\mathbf{PS}$ & -0.1739 (5e-1) & 0.3655 (1e-1) & 0.8584 (5e-6) \\
\bottomrule
\end{tabular}
}
\vspace{-10pt}
\end{table}

\begin{figure}[t]
    \centering
    \subfigure{\includegraphics[width=0.3\linewidth]{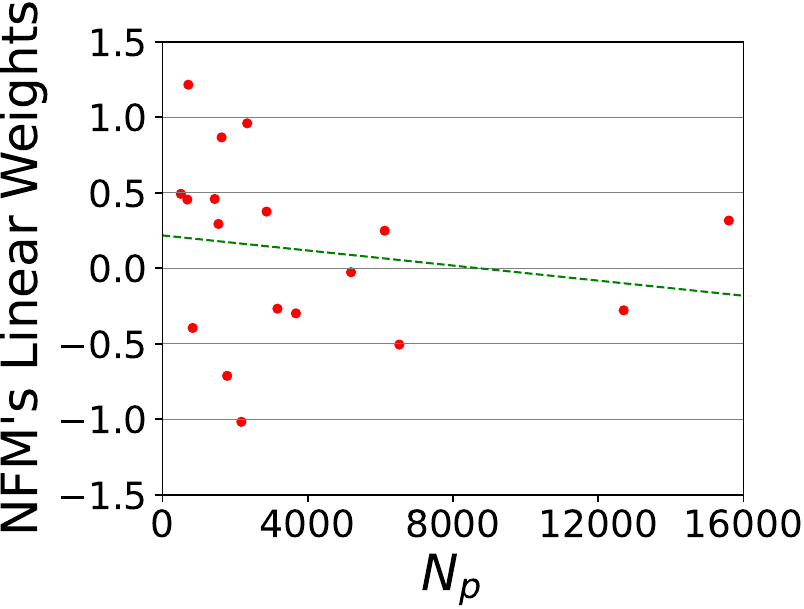}}\quad
    \subfigure{\includegraphics[width=0.3\linewidth]{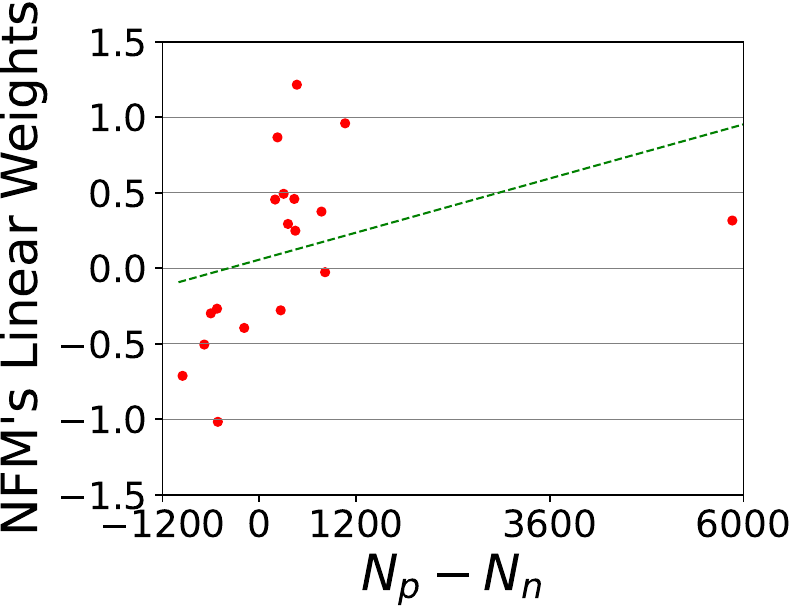}}\quad
    \subfigure{\includegraphics[width=0.3\linewidth]{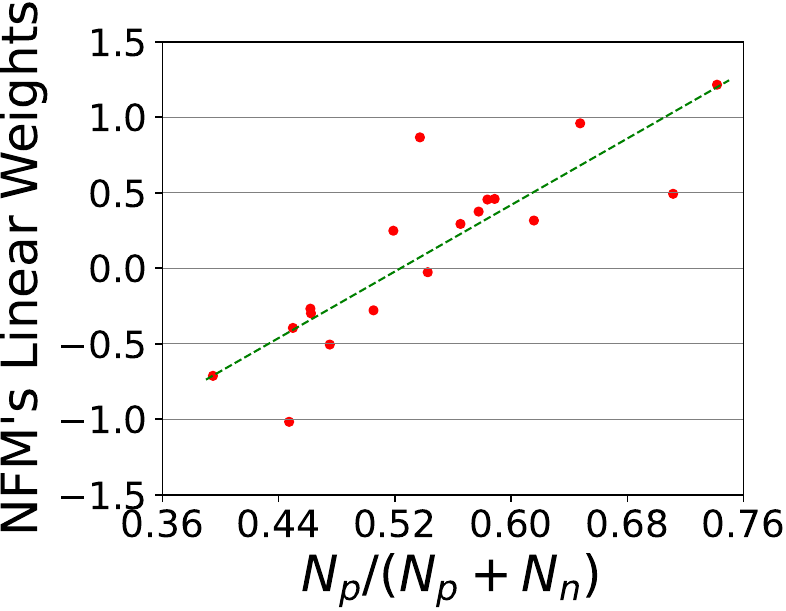}\label{fig:lw_ratio}}
    \vspace{-15pt}
    \caption{Relations between linear weights and three statistics. Each point denotes an item feature group in ML-1M and the dashed lines are obtained by linear regression.} 
    \label{fig:lw_3corr}
    \vspace{-5pt}
\end{figure}

\subsection{Statistical Analysis on Bias Cause}

\begin{figure*}
    \centering
    \subfigure[Biased positive sample ratio]{\includegraphics[width=0.225\linewidth]{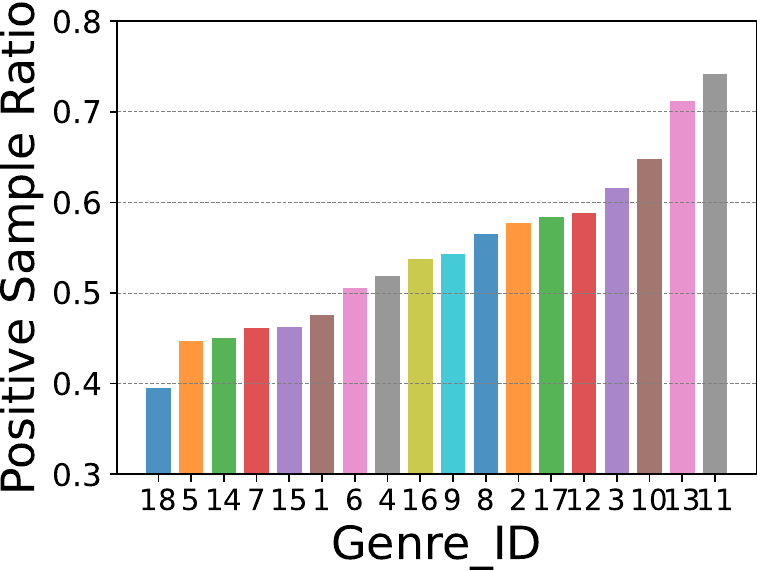}\label{subfig:chain(a)}}\quad
    \subfigure[Biased linear weights]{\includegraphics[width=0.225\linewidth]{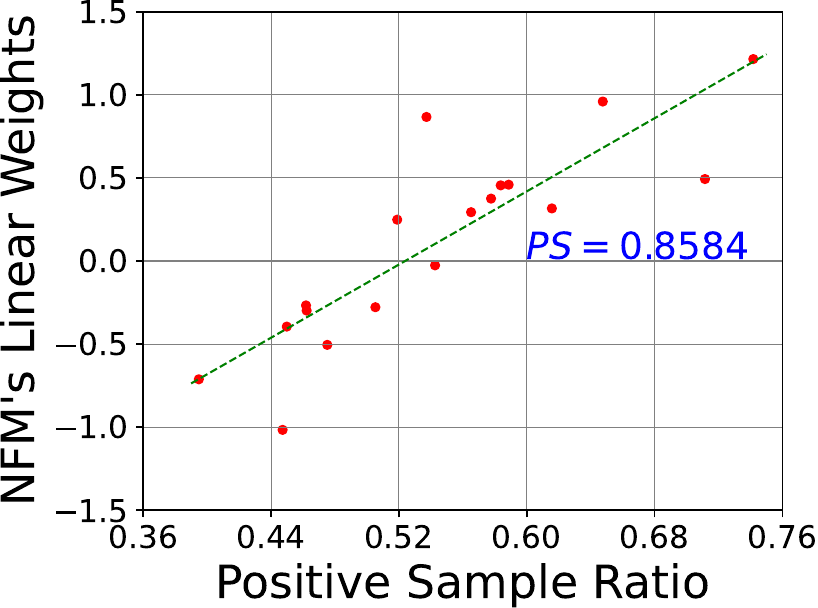}\label{subfig:chain(b)}}\quad
    \subfigure[Biased prediction scores]{\includegraphics[width=0.225\linewidth]{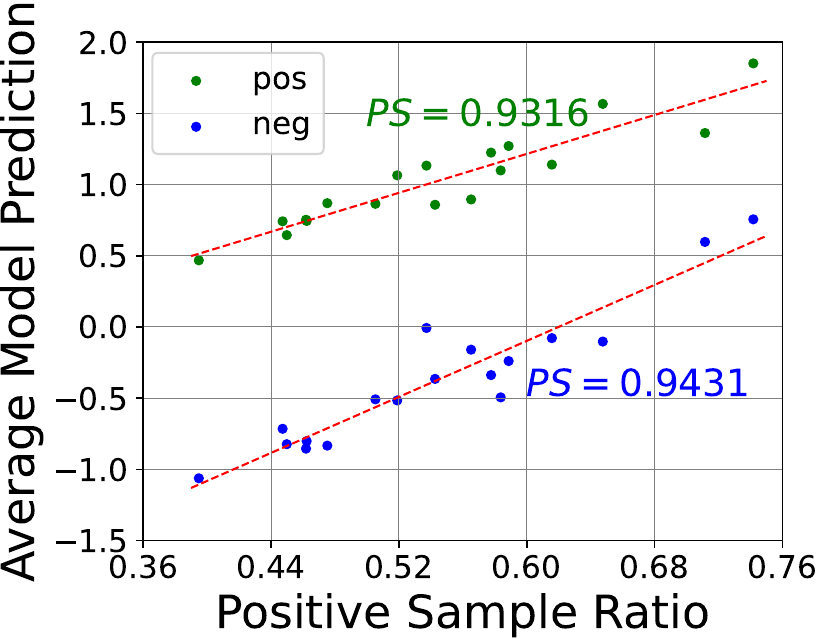}\label{subfig:chain(c)}}\quad
    \subfigure[Biased recommendations]{\includegraphics[width=0.225\linewidth]{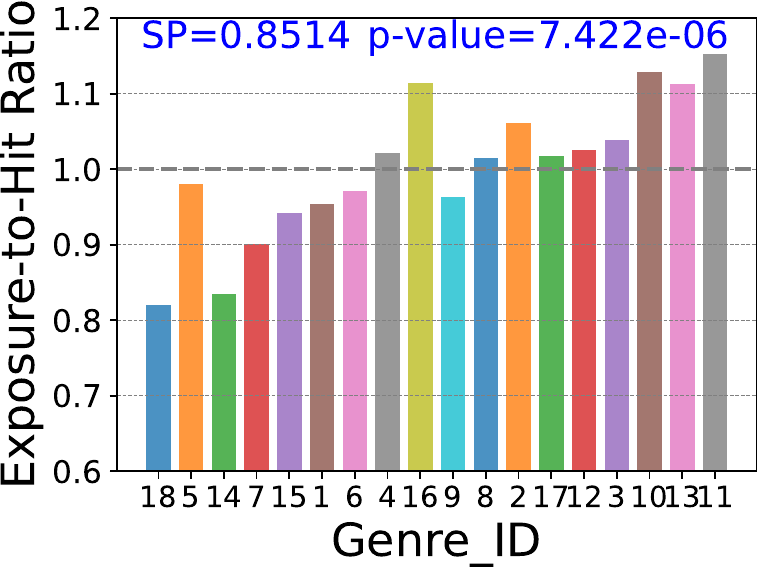}\label{subfig:chain(d)}}
    \vspace{-10pt}
    \caption{Illustration of the generation path of feature-level bias in a NFM model. In (a), each item group's positive sample ratios are sorted in ascending order. In (b)(c), we calculate the Pearson~\cite{pearson} correlation coefficients (denoted as $PS$) between the value of interested ($y$-axis) and the positive sample ratio to verify their linear relationship. In (c), "pos" and "neg" denote the average prediction scores computed over all positive or negative testing samples, respectively. In (d), $SP$ and p-value are results of Spearman test~\cite{spearman} between each item group's $EHR$ in Equation~\eqref{eq:ratio} and positive sample ratio, showing the monotonic relationship between them.
    }
    \label{fig:logical_chain}
    \vspace{-10pt}
\end{figure*}

To validate our conjecture, we investigate the relations between the linear weights of the bias field and three representative group-level statistical characteristics \wrt the bias field in the training set, including 1) {the number of positive samples in each item group} ($N_{p} \in \mathbb{R}^k$), 2) {the difference between positive and negative samples in each item group} ($N_{p}-N_{n} \in \mathbb{R}^k$), and 3) {the positive sample ratio in each item group} ($N_{p}/(N_{p}+N_{n}) \in \mathbb{R}^k$)\footnote{Note that $N_{p}/(N_{p}+N_{n})$ is a vector where the $j$-th entry denotes the positive ratio of samples with feature $x_j$.}. We take Pearson~\cite{pearson} and Spearman~\cite{spearman} tests to measure the monotonic and linear relationship between the linear weights and these statistics, respectively. The correlation coefficients with $p$-values are shown in Table~\ref{tab:twotest} and the linear weights of NFM are visualized in Figure~\ref{fig:lw_3corr}. From Figure~\ref{fig:lw_3corr} and Table~\ref{tab:twotest}, we have the following three main observations.

\begin{itemize}[leftmargin=*]
    \item The small correlation coefficients \emph{w.r.t. } $N_{p}$ show that the linear weights are not highly correlated with $N_{p}$. We thus postulate that the difference across item groups \wrt the number of positive sample is insufficient to explain the feature-level bias in the linear weights as the negative samples will decrease the weight.
    \item In Figure~\ref{fig:lw_3corr}, the feature groups with higher $N_{p}-N_{n}$ elements usually obtain higher linear weights, while the correlations break on some outliers. This is reasonable since not only the relative difference but also the absolute quantity of positive and negative samples affect the final scale of linear weights.
    \item The weights and positive sample ratios show strong linear correlations from Figure~\ref{fig:lw_ratio}, which are further verified by the large Spearman and Pearson coefficients in Table~\ref{tab:twotest} ($\text{SP}>0.87$, $\text{PS}>0.80$ with $p\text{-value}<1e\text{-}4$). The strong correlations demonstrate that the biased positive sample ratios are more likely to affect the linear weights in CTR models, further leading to the biased prediction scores and biased recommendations. 
\end{itemize}

\noindent\textbf{$\bullet$ Summary.} 
To summarize, we have discovered a generation path of feature-level bias: biased positive sample ratios $\rightarrow$ biased linear weights in CTR model $\rightarrow$ biased prediction scores $\rightarrow$ biased recommendations. Specifically, 1) as shown in Figure~\ref{subfig:chain(a)}, there exists an imbalanced distribution of the positive sample ratios across item groups. Such data bias causes feature-level bias originally\footnote{As an initial attempt, we focus on one iteration of the recommender system, leaving the consideration of feedback loop to future work.}. 2) The CTR model will learn such bias via the linear weights of the bias field. As illustrated in Figure~\ref{subfig:chain(b)}, the linear weights are highly correlate with the positive sample ratios. 3) Subsequently, the prediction score of CTR models are affected by the linear weights, causing the correlations between the average scores with the positive sample ratios as shown in Figure~\ref{subfig:chain(c)}. 4) Lastly, the feature groups with higher positive sample ratios are more likely to be over-recommended (\cf Figure~\ref{subfig:chain(d)}) and vice versa.

\section{METHODOLOGY} \label{sec:method}

According to the above analysis, the key to counteracting the feature-level bias lies in cutting off the path from the uneven distribution of positive sample ratio to the biased model predictions. Existing methods achieve the target through regularization~\cite{zhu2021popularity}, adversarial training~\cite{zhu2020measuring}, or causal inference~\cite{decrs}, which changes the learning of CTR models and inevitably brings significant overhead such as updating the system architecture to be compatible with such changes. In this light, we pursue strategies for counteracting the feature-level bias with minimally invasive changes to the recommender system. Generally, there are two options to achieve the target at the pre-training stage and the post-training stage.

\begin{itemize}[leftmargin=*]
    \item \textbf{Pre-training Adjustment.} As the uneven distribution of positive sample ratio is the start of generation path, it seems to be a promising strategy to counteract the feature-level bias by adjusting the distribution of training data. This can be easily implemented into the existing data sampler for CTR model training.
    \item \textbf{Post-training Adjustment.} As the linear weights corresponding to the bias field, \ie $\{w_j | j \in [1, k]\}$, largely explain the biased predictions of CTR models, we also consider adjusting the weights after the training of CTR models to counteract the feature-level bias. This is also minimally invasive as only changing $k$ parameters. Note that $k \ll n$ and the high-order component typically contains much more parameters than the linear part.
\end{itemize}
Undoubtedly, adjusting the data will influence the whole learning procedure, which is complex due to non-convex property of most CTR models. The effect of pre-training adjustment is thus hard to control. As such, we mainly investigate how to perform the post-training adjustment, which is formulated as:
\begin{equation}
    w_j' \longrightarrow w_j, \ j=1,\cdots,k,
\end{equation}
where $w_j'$ denotes the expected weight for the $j$-th feature. Accordingly, we can obtain the adjusted weights through the following two ways: \textit{linear weight reduction} and \textit{linear weight reconstruction}.

\subsection{Linear Weight Reduction}

To counteract the impact of biased linear weights, an intuitive operation is removing them from the biased CTR model, \ie 
\begin{equation}
    w_j' = 0, \ j=1,\cdots,k.
\end{equation}
To enhance flexibility, we seek a general form of such adjustment, which reduces linear weights with controllable strength. Formally,
\begin{equation}\small\label{eq:bias_mitigation}
    w_j'=\alpha w_j,\ j=1,\cdots,k,
\end{equation}
where $\alpha\in\left[0,1\right]$ is a specified coefficient. In particular, $\alpha=1$ indicates no adjustment and $\alpha=0$ means ignoring the linear weights in the bias field entirely. In practice, we set $\alpha=0$ for default value, which blocks the path of bias generation and reduces the feature-level bias to a large extent.

This adjustment applies very limited changes to the well trained CTR model, which should largely preserve the preference estimation since the linear weights in other fields (\eg ItemID) and the high-order part also provides sufficient information for user preference estimation. Empirical results in Table~\ref{table:nbt_performance} validate this hypothesis, where the adjusted model achieves comparable performance as the original model.

\subsection{Linear Weight Reconstruction}

\begin{figure}
    \centering
    \includegraphics[width=0.99\linewidth]{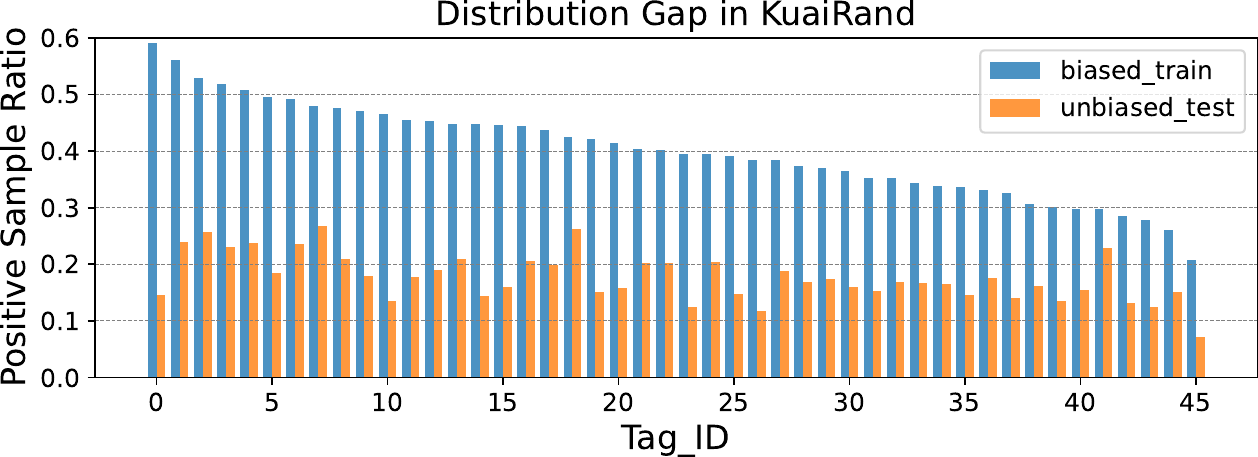}
    \vspace{-10pt}
    \caption{The distribution gap \wrt positive sample ratio between the biased data collected from the regular exposure and unbiased data collected from random exposure in the KuaiRand~\cite{kuairand} dataset.}
    \label{fig:kwairand_gap}
    \vspace{-10pt}
\end{figure}

Recall that there is a gap between the training samples collected from previous exposures (\ie non-random exposure) and the data from random exposure which reflects the true user preference. To further illustrate this phenomenon, we compute the positive sample ratios of different video tags in KuaiRand~\cite{kuairand} and visualize the distributions over both biased data from non-random exposure and unbiased data from random exposure. As shown in Figure~\ref{fig:kwairand_gap}, the biased and unbiased data exhibit huge gaps regarding the distribution of positive sample ratio. Furthermore, the positive sample ratio of the unbiased data is not uniformly distributed over bias features. As unbiased data is the golden standard for testing CTR models, we further set the target of post-training adjustment as approaching the expected linear weights as learning CTR models from the unbiased data. In this case, we assume very limited unbiased data are available as most existing works for unbiased recommendation~\cite{autodebias}. Nevertheless, biased data is still the default choice for training CTR models due to the extreme sparsity of unbiased data. Let $\mathcal{D}_u$ denote the unbiased data, we obtain the expected weights through linear weight reconstruction according to $\mathcal{D}_u$. As the target is to adjust the linear weights learned from biased data to approach those coherent with unbiased data, there are two key steps to remove the influence of positive sample ratio of the biased data and inject the influence of positive sample ratio of the unbiased data. In particular,

\textbf{Step 1: Removing the biased positive sample ratio.} As the linear weights of CTR models are highly correlated with the positive sample ratio of the biased training data, we conduct linear regression on the (ratio, weight) pairs and regress the estimated weights out to get the residuals as follows:
\begin{equation}\label{eq:regress}
    r_j=w_j-\hat{w_j},\ j=1,\cdots,k,
\end{equation}
where $\hat{w_j}$ denotes the estimation from linear regression. The residuals are no longer correlated with the positive sample ratio of the training data, thus independent from the data cause of the feature-level bias.

\textbf{Step 2: Leveraging the unbiased positive sample ratio.} We use $\mathcal{D}_u$ to estimate the ideal positive sample ratio under the random exposure, \ie the true user preference. We denote the estimated ratios as $s_1,\cdots,s_k$, and combine the ratios with the residuals linearly to construct new linear weights for the bias field:
\begin{equation}\label{eq:new_weights}
    w_j'=\beta s_j+\gamma r_j,\ j=1,\cdots,k,
\end{equation}
where $\beta>0,\gamma>0$ are hyper-parameters. We use the random exposure data $\mathcal{D}_u$ to search for optimal value of $\beta$ and $\gamma$ in practice. As such, we only estimate $k + 2$ values from $\mathcal{D}_u$, which means that the amount of required random exposure data is limited.

The overall procedures for linear weights reduction and reconstruction are shown in Algorithm~\ref{alg:lw}. It is worth mentioning that both methods do minimally invasive adjustments on the pre-trained CTR model, which are fast and energy-saving, showing promising utility in real-world scenarios.

\begin{algorithm}[t]
 \caption{Linear Weight Reduction \& Reconstruction}
 \LinesNumbered
 \label{alg:lw}
 \KwIn{A well trained CTR model $FM(\cdot)$ with biased linear weights $w_1,\cdots,w_k$; training data $\mathcal{D}$; unbiased data $\mathcal{D}_u$; hyper-parameters $\alpha,\beta,\gamma$}
 \KwOut{A debiased CTR model}

\eIf{$\mathcal{D}_u = \emptyset $}{
/* \textit{Linear Weight Reduction} */

Calculate $w_j'$ with Equation~\eqref{eq:bias_mitigation};
}
{ 
/* \textit{Linear Weight Reconstruction} */

Calculate $r_j$ with Equation~\eqref{eq:regress};

Estimate $s_j$ with $\mathcal{D}_u$;

Calculate $w_j'$ with Equation~\eqref{eq:new_weights};
}

Replace linear weights with $w_j$ = $w_j'$.

\end{algorithm}

\section{RELATED WORK} \label{sec:related}
In this section, we review recent works about \textit{bias in recommendation}, \textit{fairness in recommendation}, and \textit{CTR prediction}.

\noindent\textbf{Bias in Recommendation. }
Recommender systems utilize user-item interactions for model training. However, since the training data is observational rather than experimental~\cite{chen2020bias}, the model suffer from various biases, such as selection bias~\cite{liu2022rating}, position bias~\cite{fairstein2022external}, exposure bias~\cite{liu2020general} and popularity bias~\cite{pda,xv2022neutralizing,liu2023mitigating}. Existing works address the bias issue mainly by regularization~\cite{zhu2021popularity,abdollahpouri2017controlling,rhee2022countering}, representation learning~\cite{krishnan2018adversarial,dice,xv2022neutralizing} and causal inference~\cite{ipw,wang2019doubly,pda,wei2021model}. For example, ~\cite{zhu2021popularity} adopts the Pearson correlation coefficient as regularization to discourage the effect of popularity. ~\cite{dice} disentangles user representation into the interest and conformity part, then uses the well-trained interest embedding for debiased recommendation. ~\cite{pda} uses backdoor adjustment to block the effect of popularity on item representation and leverages popularity for better recommendation. Despite their success, they mainly focus on debiasing for collaborative filtering models~\cite{basicmf} that optimized with only interaction data. In addition to interaction data, real-world CTR models~\cite{nfm,deepfm} also rely on abundant side-information (\eg user features) for model training, making bias analysis more challenging. To tackle this challenge, our work formulates and analyzes the feature-level bias problem in CTR prediction task, and proposes simple-yet-effective strategies to counteract feature-level bias for FM-based CTR models.

\vspace{3.5pt}
\noindent\textbf{Fairness in Recommendation. }
Recently, many impressive works focus on improving fairness for recommender system~\cite{wang2023uncertainty,smith2023scoping,xu2023p}. They usually eliminate the unfairness from different perspectives (\ie user-side, item-side and multi-side). For example, \cite{li2021user,islam2021debiasing,li2021towards, rahmani2022experiments, chen2023improving} try to eliminate the unfairness between different users, \cite{beutel2019fairness,marras2022regulating,li2022fairgan,wu2023faster,chen2023fairly} dedicate to improve the fairness from item-side, while \cite{patro2020fairrec,wu2021tfrom,naghiaei2022cpfair,wu2022joint,wu2022multi,liu2023mitigating} focus on improve fairness from both sides. This work is most related to item-side fairness and equal opportunity~\cite{hardt2016equality} notion, which encourages the recommendation opportunities of different item groups to be identical. Existing methods to improve item-side fairness include adversarial training~\cite{zhu2020measuring,li2022fairgan}, regularization~\cite{beutel2019fairness,geyik2019fairness} and reinforcement learning~\cite{ge2021towards}. Consistent with previous views~\cite{li2022fairness}, we believe that bias could be the main cause of unfairness. So different from the existing works that directly optimize the various fairness metrics, we analyze the cause of feature-level bias which is responsible for unfair recommendation in CTR prediction, and improve the fairness by combating the bias.

\vspace{3.5pt}
\noindent\textbf{CTR Prediction. }
CTR prediction is one of the most important tasks in recommendation, which aims at ranking a small number of candidate items using abundant features for better personalization~\cite{huawei_BARS}. The widely-used CTR models~\cite{zhu2021open} include shallow models such as logistic regression (LR~\cite{LR}) and factorization machines (\eg FM~\cite{rendle_fm}, FFM~\cite{FFM}, HOFM~\cite{HOFM}), as well as deep models which leverages DNN (\eg Wide\&Deep~\cite{widedeep}, NFM~\cite{nfm}, DeepFM~\cite{deepfm}, xDeepFM~\cite{xdeepfm}) or attention mechanisms (\eg AFM~\cite{afm}, AutoInt~\cite{autoint}, InterHAt~\cite{InterHAt}) to model high-order feature interactions. Despite their great progress in improving recommendation accuracy, alleviating bias in CTR prediction receives little scrutiny. This work bridges the research gap by studying the feature-level bias in CTR prediction, and identifies the importance of linear part in introducing the feature-level bias.
\section{EXPERIMENTS} \label{sec:exp}

In this section, we conduct extensive experiments to answer the following three questions:
\begin{itemize}[leftmargin=*]
    \item \textbf{RQ1:} Can our proposed methods improve item-side fairness by alleviating feature-level bias?
    \item \textbf{RQ2:} How do our methods perform when predicting user preference on unbiased random exposure data? 
    \item \textbf{RQ3:} How do the designs in our methods (\eg leveraging unbiased positive ratios) affect the performance?
\end{itemize}

\subsection{Experimental Settings}

\begin{table}[t]
\caption{Statistics of the datasets. \#G is short for the number of groups. nbt and dt denote that the samples are used as normal biased and debiased validation/test data, respectively.}
\vspace{-8pt}
\label{table:dataset}
\setlength{\abovecaptionskip}{0cm}
\setlength{\belowcaptionskip}{0cm}
\setlength{\tabcolsep}{2mm}{
\resizebox{0.99\linewidth}{!}{
\begin{tabular}{ccccccccc}
\hline
Dataset  & \#User & \#Item & \#Train & \#Val-nbt & \#Val-dt & \#Test-nbt & \#Test-dt & \#G \\ \hline
\multicolumn{1}{c}{ML-1M} & 6.0k & 3.7k & 700k & 43.3k & 0 & 67.8k & 0 & 18 \\
Book & 117k & 79.2k & 3,970k & 609k & 0 & 317k & 0 & 486 \\
KuaiRand & 16.7k & 5.1k & 1,036k & 107k & 10k & 98.2k & 360k & 46 \\ \hline
\end{tabular}}}
\vspace{-10pt}
\end{table}

\subsubsection{Evaluation protocol}
 In this work, we adopt the following two offline tests~\cite{interd} to evaluate the feature-level bias.
 \begin{itemize}[leftmargin=*]
     \item \textbf{Normal biased test.} It aims at improving the click or conversion rate on user clicks collected from regular exposure. We use the linear weights reduction algorithm to improve item-side fairness with negligible sacrifice of accuracy~\cite{li2022fairness}. Accordingly, we evaluate both recommendation performance and item-side fairness under the normal biased test.
     \item \textbf{Debiased test.} 
     It evaluates recommendation performance on the unbiased random exposure data which better reflect true user preferences. As unbiased data is typically viewed as the golden standard, we focus on the recommendation performance under the debiased test.
 \end{itemize}

 \noindent \textit{Evaluation metrics.} We adopt UAUC and NDCG@K~\cite{ndcg} to evaluate the recommendation accuracy. Following~\cite{autodebias}, we calculate AUC and NDCG@K for each user and then take the average score of all users to obtain UAUC and NDCG@K. As to the item-side fairness, we follow the previous work~\cite{zhu2020measuring} to compute the relative standard deviation of successful recommendation probabilities regarding users' top-$K$ list (\ie $P(j,K)$) over item groups. Formally,
 \begin{equation}\label{eq:reo}
    REO@K=\frac{
        \mathcal{S}\left(
            \left\{
                P\left(
                    j, K
                \right) | j \in [1, k]
            \right\}
        \right)
    }{\frac{1}{k} \sum_{j \in [1, k]} P\left(
        j, K
    \right)},
\end{equation}
where $\mathcal{S}(\cdot)$ calculates standard deviation. Apparently, the value of $REO@K$ is in the range of $[0, +\infty)$, and smaller value of $REO@K$ indicates less feature-level bias and better item-side fairness. $REO@K=0$ means the absolute equal opportunity~\cite{hardt2016equality} is achieved. In the experiment we take $K=5$ as default.

\begin{table*}[t]
\vspace{-15pt}
\setlength{\abovecaptionskip}{0cm}
\setlength{\belowcaptionskip}{0cm}
\caption{Performance comparison on the normal biased tests. The best and second best results are highlighted in bold and underlined, respectively. \%improve denotes the relative performance improvements of our method compared to the basemodel.}
\label{table:nbt_performance}
\begin{center}
\setlength{\tabcolsep}{0.9mm}{
\resizebox{\textwidth}{!}{
\begin{tabular}{l|ccc|ccc|ccc|ccc|ccc|ccc}
\toprule
\textbf{} & \multicolumn{9}{c|}{\textbf{FM}} & \multicolumn{9}{c}{\textbf{NFM}} \\ \hline 
\multirow{2}{*}{\textbf{Method}} & \multicolumn{3}{c|}{\textbf{ML-1M}} & \multicolumn{3}{c|}{\textbf{Book}} & \multicolumn{3}{c|}{\textbf{KuaiRand}} & \multicolumn{3}{c|}{\textbf{ML-1M}} & \multicolumn{3}{c|}{\textbf{Book}} & \multicolumn{3}{c}{\textbf{KuaiRand}} \\
 & \textbf{UAUC$\uparrow$} & \textbf{NDCG$\uparrow$} & \textbf{REO$\downarrow$} & \textbf{UAUC$\uparrow$} & \textbf{NDCG$\uparrow$} & \textbf{REO$\downarrow$} & \textbf{UAUC$\uparrow$} & \textbf{NDCG$\uparrow$} & \textbf{REO$\downarrow$} & \textbf{UAUC$\uparrow$} & \textbf{NDCG$\uparrow$} & \textbf{REO$\downarrow$} & \textbf{UAUC$\uparrow$} & \textbf{NDCG$\uparrow$} & \textbf{REO$\downarrow$} & \textbf{UAUC$\uparrow$} & \textbf{NDCG$\uparrow$} & \textbf{REO$\downarrow$}  \\ \hline 
\textbf{Basemodel} & \textbf{0.7266} & \textbf{0.8234} & 0.5577 & \underline{0.6229} & \textbf{0.8480} & 0.1624 & \underline{0.6337} & \underline{0.7260} & 0.3857 & 0.7272 & 0.8298 & 0.4969 & 0.6211 & 0.8475 & 0.1620 & 0.6359 & 0.7278 & 0.4036 \\ \hline
\textbf{Unawareness} & 0.7210 & \underline{0.8233} & 0.4814 & 0.6169 & 0.8465 & \underline{0.1421} & 0.6286 & 0.7238 & 0.4025 & \textbf{0.7285} & \underline{0.8324} & 0.5103 & 0.6218 & 0.8467 & \underline{0.1563} & \textbf{0.6367} & \textbf{0.7286} & 0.3959 \\
\textbf{w/o l/p} & 0.7100 & 0.8095 & 0.7267 & 0.6205 & 0.8464 & 0.1551 & 0.6234 & 0.7191 & 0.4002 & 0.7151 & 0.8197 & 0.6294 & 0.6193 & 0.8463 & 0.1650 & 0.6231 & 0.7187 & 0.3699 \\
\textbf{w/o b/l/w} & 0.7193 & 0.8145 & 0.5507 & 0.6204 & 0.8463 & 0.1549 & 0.6252 & 0.7206 & 0.4342 & 0.7224 & 0.8248 & 0.5229 & 0.6219 & 0.8467 & 0.1558 & \underline{0.6367} & \underline{0.7285} & 0.3959 \\
\textbf{IPW}~\cite{ipw} & 0.7112 & 0.8145 & \underline{0.3537} & 0.6209 & 0.8472 & 0.1648 & 0.6052 & 0.7051 & \underline{0.2925} & 0.7233 & 0.8307 & \underline{0.3857} & 0.6208 & 0.8465 & 0.1667 & 0.6144 & 0.7115 & \underline{0.3114} \\
\textbf{DecRS}~\cite{decrs} & 0.7202 & 0.8159 & 0.4350 & \textbf{0.6230} & \underline{0.8478} & 0.1596 & 0.6267 & 0.7212 & 0.4225 & \underline{0.7282} & \textbf{0.8349} & 0.5158 & \textbf{0.6239} & \textbf{0.8481} & 0.1627 & 0.6344 & 0.7265 & 0.3941 \\ 
\textbf{FairGo}~\cite{fairgo} & \underline{0.7248} & 0.8152 & 0.5230 & 0.6222 & 0.8474 & 0.1638 & \textbf{0.6337} & \textbf{0.7263} & 0.3899 & 0.7246 & 0.8238 & 0.5253 & \underline{0.6234} & \underline{0.8481} & 0.1851 & 0.6288 & 0.7239 & 0.4273 \\ \hline
\textbf{Reduction} & 0.7172 & 0.8153 & \textbf{0.2956} & 0.6190 & 0.8461 & \textbf{0.1391} & 0.6233 & 0.7193 & \textbf{0.2829} & 0.7167 & 0.8139 & \textbf{0.3351} & 0.6171 & 0.8458 & \textbf{0.1341} & 0.6243 & 0.7208 & \textbf{0.3058} \\
\textbf{\%improve} & -1.29\% & -0.98\% & 47.00\% & -0.63\% & -0.22\% & 14.35\% & -1.64\% & -0.92\% & 26.55\% & -1.44\% & -1.92\% & 32.56\% & -0.64\% & -0.20\% & 17.22\% & -1.82\% & -0.96\% & 24.23\% \\
\bottomrule
\end{tabular}
}}
\end{center}
\vspace{-10pt}
\end{table*}

\subsubsection{Datasets}
We conduct experiments on the following three datasets: (1) \textbf{ML-1M}\footnote{\url{https://grouplens.org/datasets/movielens/1m/}.} is a popular movie recommendation dataset which involves abundant features (\eg user gender, age, occupation and movie genre). (2) \textbf{Book} is one of the Amazon product datasets\footnote{\url{https://jmcauley.ucsd.edu/data/amazon/}.} with book categories. (3) \textbf{KuaiRand}~\cite{kuairand} is a short-video recommendation dataset on the video-sharing platform Kuaishou\footnote{\url{https://www.kuaishou.com}.} which contains interactions from both normal and random exposure. We sort and split the interactions into training, validation, and test set chronologically. For ML-1M and Book, interactions with rating value larger than $3$ are treated as positive samples otherwise negative. For KuaiRand, we treat the interaction with $is\_click=1$ as positive samples otherwise negative. Considering that random exposure data is usually at the high expense of hurting user experience~\cite{schnabel2016recommendations,autodebias}, we randomly sample $10,000$ interactions from the unbiased validation data. For each dataset, we adopt the 20-core filtering setting and split the items into groups based on movie genre, book category, and video tag respectively. The statistics of these datasets are summarized in Table~\ref{table:dataset}.

\subsubsection{Baselines}
We select two representative CTR models, FM~\cite{rendle_fm} and NFM~\cite{nfm}, as our backbone models. On \textit{normal biased tests}, we select the following heuristic, debias, and fairness methods which may alleviate the bias amplification in CTR prediction tasks:
\begin{itemize}[leftmargin=*]
    \item \textbf{Unawareness} removes the bias field (\eg movie genres in ML-1M) from the data and leverages the remaining fields for training.
    \item \textbf{w/o l/p} is a variant of FM/NFM which only keeps the high-order part (\ie $f_{\theta}(\bm{x})$ in Equation~\eqref{eq:fm_form}) for training.
    \item \textbf{w/o b/l/w} removes linear weights in FM/NFM about the bias field and keeps the remaining parts for training. In contrast, our method trains a whole CTR model and adjust the linear weights in the inference stage.
    \item \textbf{IPW}~\cite{ipw} is the inverse propensity weighting method in causal debiasing. We use each group’s positive sample ratio as propensity scores to down-weight the items in advantaged groups. 
    \item \textbf{DecRS}~\cite{decrs} uses causal inference technique to alleviate the bias amplification issue for CTR models. We implement the DecRS on FM and NFM backbones.
    \item \textbf{FairGo}~\cite{fairgo} is a method that pursues fair recommendation by removing the information of sensitive features through adversarial training. We implement FairGo on FM and NFM by regarding the features in the bias field as sensitive features. We tune the bias control coefficient $\lambda$ in [1e-3,10] at 5x multiplicative ratio.
\end{itemize}
On \textit{debiased tests}, as the linear weight reconstruction algorithm leverages a few random exposure records, we further compare two baselines which also use random exposure data:
\begin{itemize}[leftmargin=*]
    \item \textbf{Finetune} is a heuristic method that finetunes the pre-trained FM/NFM model on unbiased validation data.
    \item \textbf{InterD}~\cite{interd} is a multi-teacher distillation method that aims to achieve better performances on both tests. We adopt the pre-trained biased model as the biased-teacher and the model finetuned on unbiased validation data as the unbiased-teacher. The student model is another NFM/FM which has the same settings as both teachers.
\end{itemize}

\subsubsection{Hyper-parameter settings}
We take the same experimental setups for all the baselines to compare them fairly. The FM and NFM are implemented with the popular CTR benchmark DeepCTR\footnote{\url{https://github.com/shenweichen/DeepCTR-Torch}.}. All the baselines are optimized by the Adam~\cite{adam} optimizer with batch size 256 and BCE loss. For FM, we tune the learning rate in $\{5e-4,1e-3,5e-3\}$. For NFM, we additionally tune the dropout rate of DNN and bi-interaction module in $\{0.1,0.2,0.5\}$. We also tune the regularization coefficients in $\{1e-6,5e-6,1e-5,1e-4,1e-3\}$. For the linear weight reduction algorithm, we set $\alpha=0$ by default. For the linear weight reconstruction algorithm, we search both $\beta$ and $\gamma$ in $\{1,2,4,8,12,16,20\}$.

\subsection{Performance Comparison}

\subsubsection{Improving item-side fairness on normal biased tests (RQ1)}\label{sec:nbt_analysis} Table ~\ref{table:nbt_performance} reports the recommendation accuracy and fairness metrics on normal biased tests of three datasets. From the two tables, we have the following main observations:
\begin{itemize}[leftmargin=*]
    \item On three datasets, our method (\ie linear weight reduction) remarkably reduces REO@5 with only little sacrifice of accuracy. Specifically, the unfairness decrease of our method over the basemodel \wrt REO@5 is from 14.35\% to 47.00\% on different datasets,  which shows that our method significantly improves item-side fairness. Meanwhile, the average performance drop is negligible \wrt UAUC and NDCG@5 (-1.06\% on average across all cases).
    \item IPW and Unawareness can reduce REO@5, but their effectiveness is sensitive to datasets (\eg IPW fails on Book). IPW assigns a propensity weight for each training sample, affecting both linear and high-order weights. Our method directly post-adjusts linear weights, which is thus more robust and can alleviate bias consistently. This is the benefit of recognizing the role of CTR model components \wrt the feature-level bias. As for Unawareness, we postulate that CTR models can implicitly infer information in the bias field~\cite{zhao2022towards} from the remaining features and historical interactions. Consequently, the linear part is still highly correlated to the biased positive ratios. 
    \item As two variants of linear weight reduction algorithm, w/o l/p and w/o b/l/w encounter performance degradation in most cases; meanwhile, they cannot remarkably improve item-side fairness. This indicates that isolating the linear part in the training stage cannot solve the feature-level bias effectively. As such, this result further validate the rationality of post-training adjustment, \ie adjusting the linear weights of trained CTR models.
    \item Although DecRS and FairGo sacrifice minor accuracy in some cases, they cannot reduce REO@5 effectively in most cases. The main reason is that they alleviate biases by improving high-order embedding (DecRS introduces confounded embedding for users and tags and FairGo removes the sensitive information from embeddings), whereas the linear part is more important in introducing feature-level bias.
\end{itemize}

\subsubsection{Improving accuracy on debiased tests (RQ2)}
We present the performance on the debiased test of KuaiRand in Table~\ref{table:dt_performance}, where we have the following findings:

\begin{table}[t]
\vspace{-20pt}
\setlength{\abovecaptionskip}{0cm}
\setlength{\belowcaptionskip}{0cm}
\caption{Performance on the debiased test of KuaiRand. The best and second best results are highlighted in bold and underlined, respectively. \%improve denotes the relative performance improvements of our method compared to the basemodel. Debiased tests are unavailable in ML-1M and Book.}
\label{table:dt_performance}
\setlength{\tabcolsep}{2mm}{
\resizebox{0.99\linewidth}{!}{
\begin{tabular}{lcccc}
\toprule
\multirow{2}{*}{\textbf{Method}} & \multicolumn{2}{c}{\textbf{FM}} & \multicolumn{2}{c}{\textbf{NFM}} \\
 & \textbf{UAUC$\uparrow$} & \textbf{NDCG@5$\uparrow$} & \textbf{UAUC$\uparrow$} & \textbf{NDCG@5$\uparrow$}\\ \hline 
\textbf{Basemodel} & 0.5748 & 0.4151 & 0.5735 & 0.4140 \\ \hline
\textbf{Unawareness} & 0.5655 & 0.4199 & 0.5743 & 0.4133 \\
\textbf{w/o l/p} & 0.5702 & 0.4126 & 0.5700 & 0.4175 \\
\textbf{w/o b/l/w} & 0.5706 & 0.4124 & 0.5743 & 0.4133 \\
\textbf{IPW}~\cite{ipw} & 0.5666 & 0.4144 & 0.5618 & 0.4112 \\
\textbf{DecRS}~\cite{decrs} & 0.5717 & 0.4127 & 0.5745 & 0.4155 \\ 
\textbf{FairGo}~\cite{fairgo} & 0.5745 & 0.4127 & 0.5753 & 0.4156 \\
\textbf{Finetune} & \underline{0.5795} & \underline{0.4238} & \underline{0.5933} & \underline{0.4427} \\
\textbf{InterD}~\cite{interd} & 0.5777 & 0.4191 & 0.5852 & 0.4320 \\ \hline
\textbf{Reconstruction} & \textbf{0.5999} & \textbf{0.4465} & \textbf{0.6003} & \textbf{0.4465} \\
\textbf{\%improve} & 4.37\% & 7.56\% & 4.67\% & 7.85\% \\
\bottomrule
\end{tabular}
}
}
\vspace{-10pt}
\end{table}

\begin{itemize}[leftmargin=*]
    \item Our method (\ie linear weight reconstruction) achieves the best performance among all baselines, validating the superiority of our method. Specifically, the average relative improvements over the basemodel \wrt UAUC and NDCG@5 are 4.52\% and 7.71\%, respectively. We attribute this superior performance to using estimated positive sample ratios to construct new linear weights, which can better reflect the true user preference. 
    \item All the baselines without leveraging random exposure data (Unawareness, w/o l/p, w/o b/l/w, IPW, DecRS, FairGo) show inferior results than Finetune, InterD and our method on the \textit{debiased test}. This demonstrates the effectiveness of using random exposure data for estimating the actual user preference. Nevertheless, our method still surpasses Finetune and InterD, since the linear weight reconstruction avoids the impact of data sparsity.
\end{itemize}

\subsection{In-depth Analysis (RQ3)}
To further explore what influences the effectiveness of our methods, we conduct the following parameter analysis and ablation studies.

\begin{figure}
    \centering
    \subfigure[\textbf{ML-1M}]{\includegraphics[width=0.3\linewidth]{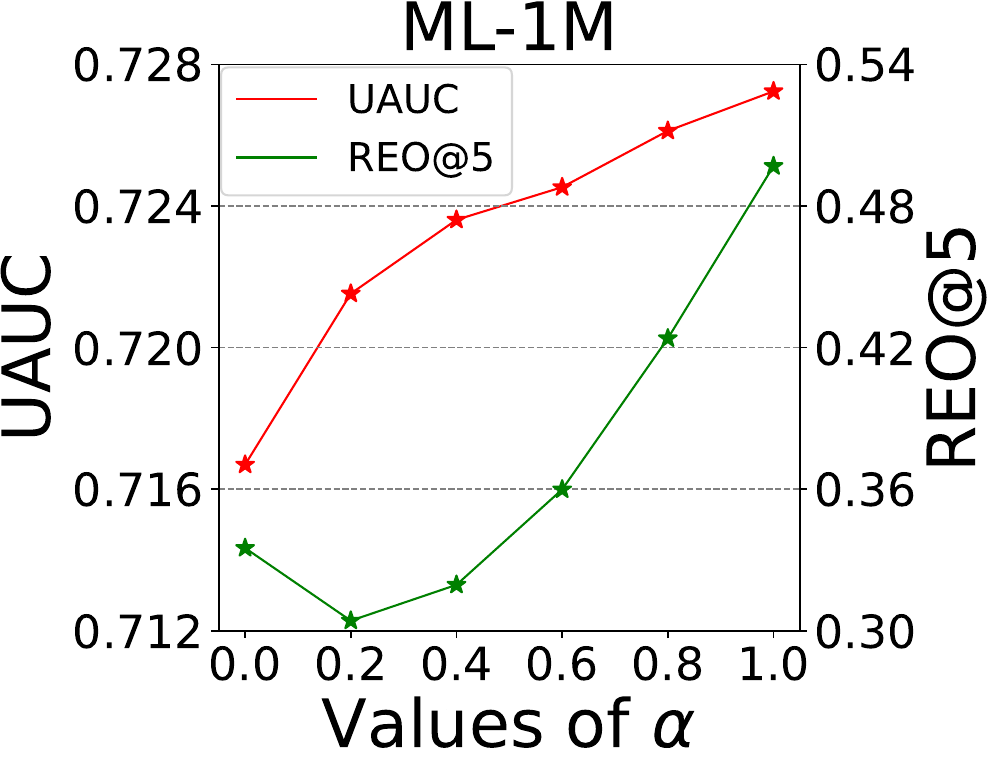}}\quad
    \subfigure[\textbf{Book}]{\includegraphics[width=0.3\linewidth]{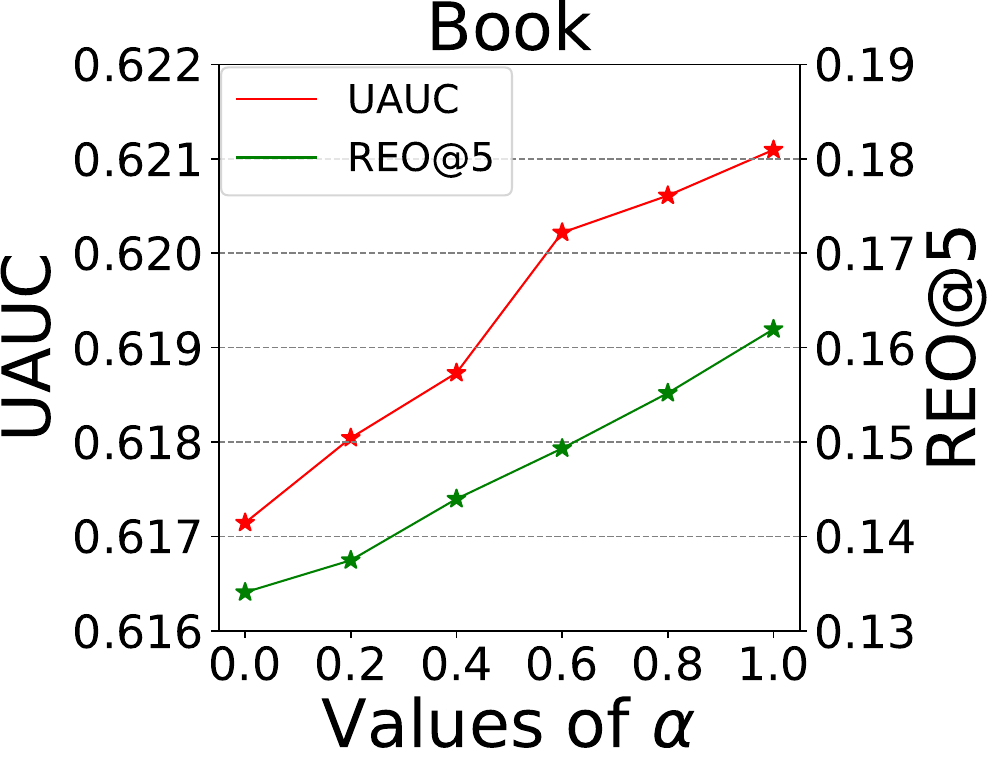}}\quad
    \subfigure[\textbf{KuaiRand}]{\includegraphics[width=0.3\linewidth]{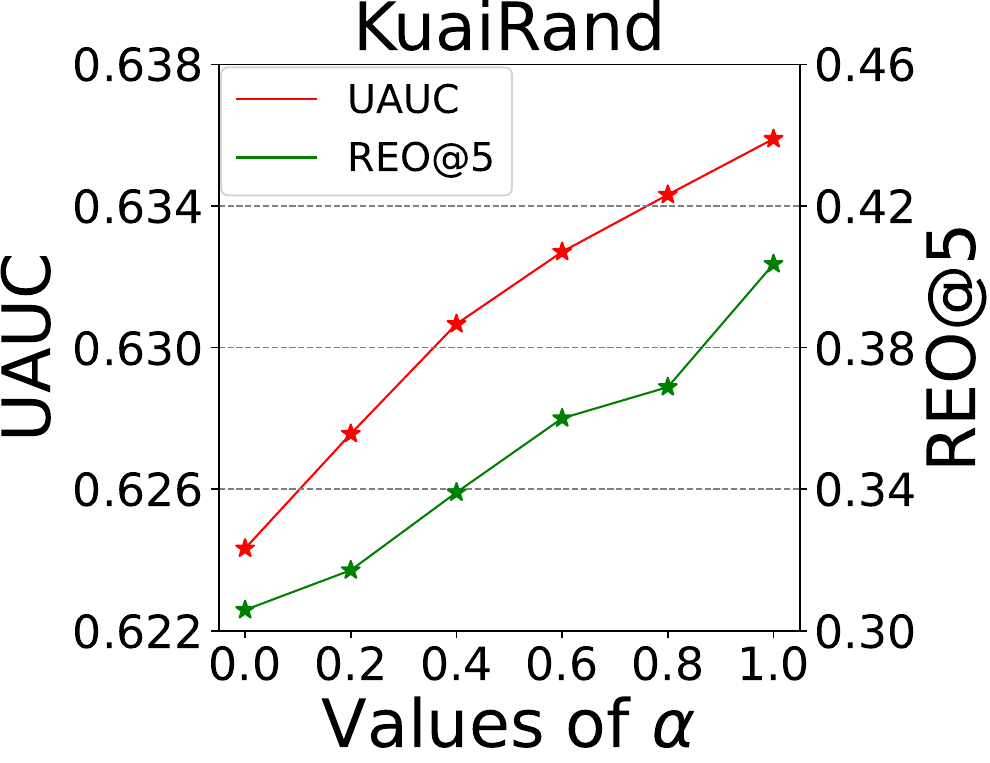}}
    \vspace{-10pt}
    \caption{Fairness-accuracy trade-off of NFM as $\alpha$ changes in Equation~\eqref{eq:bias_mitigation} on the normal biased tests.}
    \label{fig:tradeoff}
    \vspace{-10pt}
\end{figure}

\subsubsection{Study of $\alpha$ in Equation ~\eqref{eq:bias_mitigation}}\label{sec:tradeoff}
In previous experiments to improve item-side fairness on normal biased tests, we set 0 as the default value of $\alpha$ in Equation ~\eqref{eq:bias_mitigation}. In practice, the platform-side can set $\alpha\in\left[0,1\right]$ to control the fairness-accuracy trade-off as shown in Figure~\ref{fig:tradeoff}. There are several findings: (1) As $\alpha$ increases, the UAUC of different datasets keep increasing, demonstrating that the linear weights of the bias field do contribute to recommendation accuracy. The more linear weights keep, the less accuracy is sacrificed. (2) On Book and KuaiRand, REO@5 keeps increasing with an increasing value of $\alpha$, which demonstrates the fairness-accuracy trade-off phenomenon. However, on ML-1M, REO@5 firstly decreases and then increases, which suggests that when $\alpha$ is small, there exists a win-win solution in improving both fairness and accuracy. This sheds light on the choice of $\alpha$. Specifically, we recommend to select small but nonzero values (\eg $\alpha=0.1,0.2$), which can always alleviate item-side unfairness greatly, and may further approach pareto optimality~\cite{pareto1971manual} in some lucky cases.

\subsubsection{Ablation study}\label{sec:abalation} In our proposed linear weight reconstruction algorithm which aims at improving performances on debiased tests, there are mainly two parts---residual part $r_i$ and positive ratio part $s_i$. To check their effectiveness, we compare our method with the following three variants. \textit{w/ finetuning LW} denotes initializing the FM/NFM with the pre-trained biased model and only finetuning linear weights of the bias field with the unbiased validation data. \textit{w/o ratio} denotes that only using $r_j$ to construct weights (\ie $\beta=0$ in Equation~\eqref{eq:new_weights}), while \textit{w/o residual} only uses $s_j$ for construction (\ie $\gamma=0$ in Equation~\eqref{eq:new_weights}).

From results in Table~\ref{table:variants}, there are three key observations. (1) The vanilla method consistently outperforms \textit{w/o ratio} and \textit{w/o residual}. In addition, both variants still outperform the basemodel. This demonstrates the effectiveness of each part of our method. (2) It is worth noticing that although \textit{w/o ratio} doesn't require random exposure data, it outperforms the basemodel, showing potential use in cases without unbiased data. (3) Although \textit{w/ finetuning LW} can also improve performance, it doesn't outperform our construction method, which shows the limitation of simply finetuning linear weights of the bias field.

\begin{table}[t]
\vspace{-20pt}
\setlength{\abovecaptionskip}{0cm}
\setlength{\belowcaptionskip}{0cm}
\caption{Performance comparison between different variants of our method on the debiased test. The best and second best results are highlighted in bold and underlined, respectively.}
\label{table:variants}
\setlength{\tabcolsep}{2mm}{
\resizebox{0.99\linewidth}{!}{
\begin{tabular}{lcccc}
\toprule
\multirow{2}{*}{\textbf{Method}} & \multicolumn{2}{c}{\textbf{FM}} & \multicolumn{2}{c}{\textbf{NFM}} \\
 & \textbf{UAUC} & \textbf{NDCG@5} & \textbf{UAUC} & \textbf{NDCG@5}\\ \hline 
\textbf{Basemodel} & 0.5748 & 0.4151 & 0.5735 & 0.4140 \\ 
\textbf{Reconstruction} & \textbf{0.5999} & \textbf{0.4465} & \textbf{0.6003} & \textbf{0.4465} \\
\quad\textbf{w/ finetuning LW} & 0.5749 & 0.4153 & 0.5928 & 0.4392 \\
\quad\textbf{w/o ratio} & 0.5865 & 0.4304 & 0.5791 & 0.4197 \\
\quad\textbf{w/o residual} & \underline{0.5952} & \underline{0.4427} & \underline{0.5959} & \underline{0.4436} \\ 
\bottomrule
\end{tabular}
}
}
\vspace{-10pt}
\end{table}
\section{CONCLUSION AND FUTURE WORK}
In this work, we studied the feature-level bias issue in CTR prediction task, and discussed its negative impacts on item-side fairness and user experiences. We then conducted top-down analysis on CTR models, recognized the key influence of the model's linear part on feature-level bias, where positive sample ratios of the training data significantly correlates with linear weights. Upon this understanding, we propose two simple-yet-effective strategies to counteract the feature-level bias, whose key idea lies in adjusting biased linear weights to cut off the bias generation path. Our method is minimally invasive to CTR models, whose effectiveness is validated by extensive experiments on both biased and debiased tests of real-world datasets.

In the future, we would like to explore how high-order bias (\ie biases related to multi-features) is generated inside the model and the approach to addressing the issue without retraining. Lastly, our work is limited to debiasing classical FM-based CTR models with linear part, thus how to extend our analysis and method to more general and complicated CTR models (\eg DIN~\cite{DIN}) is an interesting topic. Besides, it is valuable to explore the transferability of linear weights across CTR models (\eg whether transfering unbiased linear weights from a debiased model to another biased CTR model will help alleviate feature-level bias or not).

\clearpage
\bibliographystyle{ACM-Reference-Format}
\bibliography{sample-base}

\end{document}